\documentclass[aps,prb,twocolumn,superscriptaddress,showpacs,floatfix]{revtex4}
\usepackage{epstopdf}
\usepackage{graphicx}
\usepackage{dcolumn}
\usepackage{bm}
\usepackage{amsfonts}	
\usepackage{amsmath}
\usepackage{color}

\begin{document}
\renewcommand{\thefigure}{\arabic{figure}}
\def\be{\begin{equation}}
\def\ee{\end{equation}}
\def\ber{\begin{eqnarray}}
\def\eer{\end{eqnarray}}

\def\kv{{\bf k}}
\def\qv{{\bf q}}
\def\pv{{\bf p}}
\def\sigmav{{\bf \sigma}}
\def\tauv{{\bf \tau}}
\newcommand{\h}[1]{{\hat {#1}}}
\newcommand{\hdg}[1]{{\hat {#1}^\dagger}}
\newcommand{\bra}[1]{\left\langle{#1}\right|}
\newcommand{\ket}[1]{\left|{#1}\right\rangle}

\title{ Indirect exchange interaction between magnetic adatoms in a monolayer MoS$_2$}
\date{\today}

\author{F. Parhizgar}
\affiliation{School of Physics, Institute for Research in
Fundamental Sciences (IPM), Tehran 19395-5531, Iran}
\author{H. Rostami}
\affiliation{School of Physics, Institute for Research in
Fundamental Sciences (IPM), Tehran 19395-5531, Iran}
\author{Reza Asgari}
\email{asgari@ipm.ir} \affiliation{School of Physics, Institute
for Research in Fundamental Sciences (IPM), Tehran 19395-5531,
Iran}

\begin{abstract}
We study the Ruderman-Kittle-Kasuya-Yosida (RKKY) interaction in
a monolayer MoS$_2$. We show that the rotation of the itinerant electron spin
due to the spin-orbit coupling causes a twisted interaction between
two magnetic adatoms which consists of different RKKY coupling terms,
the Heisenberg, Dzyaloshinsky-Moriya and Ising interactions. We find that
the interaction terms are very sensitive to the Fermi energy values and
change dramatically from doped to undoped systems.
A finite doping causes that all parts of the interaction
oscillate with the distance of two magnetic impurities, $R$ and the
interaction behaves like $R^{-2}$ for the long distance between two localized spins. We explore a beating pattern of
oscillations of the RKKY interaction which occurs for the doped system.
\end{abstract}
\pacs{75.30.Hx, 75.10.Lp, 73.63.-b} \maketitle
%75.30.Hx: magnetic impurity interactions
%Magnetic ordering: band and itinerant models, 75.10.Lp

\section{Introduction}

Two dimensional (2D) materials can be mostly exfoliated into individual thin layers from stacks of strongly bonded layers with weak interlayer interaction. A famous example is graphene~\cite{novoselov} and its analogs such as boron nitride~\cite{dean}.
The 2D exfoliate versions of transition metal dichalcogenides~\cite{mattheis} exhibit properties that are complementary to distinct from those in graphene. MoS$_2$ ( the polytype is 2H-MoS$_2$ has trigonal prismatic coordination) is a hexagonal crystal layered structure with a covalently bonded S-Mo-S hexagonal quasi-two dimensional network which does not have inversion symmetry, packed by weak van der Waals interactions. The monolayer of MoS$_2$ has provided a new material with a peculiar structure for the charge and the spin interactions~\cite{wang}. There is a transition from an indirect band gap of $1.3$ eV in a bulk structure to a direct band gap of $1.8$ eV in the monolayer structure~\cite{kuc}. This direct band gap opens the possibility of many optoelectronic applications~\cite{kuc}. The electronic structure of MoS$_2$ also enables valley polarization~\cite{mak, zeng, cao, splendiani} since both the conduction and valence band edges have two energy-degenerate valleys at the corners of the first Brillouin zone and valley selective circular dichroism arising from its symmetry.

Ruderman-Kittle-Kasuya-Yosida (RKKY) interaction is an indirect interaction between nuclei of transition metals or magnetic impurities, mediated by the conduction electrons~\cite{ref:RK, ref:Kasuya, ref:Yosida}. This interaction is directly related to the spin susceptibility of the host metal.
Rudermann and Kittel first suggested~\cite{ref:RK} that the spin oscillatory
interaction in metals could provide a long-range interaction
between nuclear spins and explored the effective magnetic interaction between nuclei of transition metals and thus described the reason for broadening of spin resonance line diagram. Due to the fact that the RKKY interaction is originated by the
exchange coupling between the impurity moments and the spin of
itinerant electrons in the bulk of the system, a spin splitting owing to spin-orbit coupling, must have opposite sign at the two valleys by the requirement of time-reversal symmetry, is expected to influence directly this interaction. Since the two inequivalent valleys in the monolayer MoS$_2$ are separated in the Brillouin zone by a large momentum, in the case of the absence of short-range interactions, intervalley scattering~\cite{lu} should be negligible and thus the valley index becomes a new quantum number. Therefore, manipulating valley quantum number can produce new physical effects.

In the monolayer MoS$_2$, the highest energy valence bands and the lowest energy conduction bands, at the $K$ and $K'$ valleys in momentum space, are mainly of molybdenum $d-$orbital character and spin-orbit interactions split~\cite{xiao} the valence bands by $2\lambda\sim 150$ meV leading to a spin polarization of the valence band. A minimal effective band model Hamiltonian of MoS$_2$ is found by Xiao et al~\cite{xiao} with parameters based on first principle calculations. To first order in momenta, the first part of Hamiltonian describes the dynamics of massive Dirac fermions which are studied from the graphene committees. It is commonly known that the RKKY interaction in massless graphene is quite different~\cite{ref:all, ref:sherafati, parhizgar} from that of a shcr\"{o}dinger 2D electron liquid. The second part of the Hamiltonian illustrates the spin-orbit interaction which leads to coupled spin and valley physics in the monolayer MoS$_2$. The RKKY interaction usually yields a Heisenberg coupling refereing to a parallel or antiparallel coupling of localized spins. If spin of conduction electrons processes, it can be possibly to produce a noncollinear Dzyaloshinsky-Moriaya ( DM) coupling of localized spins. We thus use this model Hamiltonian to calculate the exchange interaction between localized spins on the MoS$_2$ and show that, in particular, combination of the
spin-valley with a massive Dirac-like spectrum can mediate a much
richer collective behavior of magnetic adatoms.

The RKKY interaction in 2D electron gas in the presence of a Rashba spin-orbit coupling has been studied by Imamura et al.~\cite{ref:Imamura} and they showed that rotation of the spin of conduction electrons causes a twisted RKKY interaction which consists of three different terms, the Heisenberg, DM and Ising interactions. Moreover, the RKKY interaction between localized magnetic moments in a disordered 2D electron gas with both Rashba and Dresselhaus spin-orbit couplings has been studied~\cite{chesi}. The disorder-averaged susceptibility leads to a twisted exchange interaction suppressed exponentially with distance, whereas the second-order correlations, which determine the fluctuations of the RKKY energy, decay with the same power law as in the clean case.

In this work, we calculate the RKKY interaction mediated by
spin-orbit coupled massive Dirac fermions in a monolayer MoS$_2$ using the
Green's function method. We show that the interaction consists of three terms without having the Rashba or Dresselhaus interaction and show particularly that the
interaction behaves like $R^{-2}$ for the long distance between two localized spins. The DM coupling leads to a rotation of the spinors of magnetic moments with respect to each other. In addition a beating pattern for the interaction, in the cases where the monolayer MoS$_2$ system is doped, is obtained. As we mentioned before, the model Hamiltonian can be reduced to a massive graphene by discarding the spin-orbit coupling. Besides $R^{-3}$ dependence of the interaction for undoped massless graphene, we find that the interaction decays like $R^{-2}$ for a massive and doped graphene. However, for a massive and undoped graphene, the interaction decays rapidly in a good agreement with those results obtained in Ref.~[\onlinecite{dugeav}].

The paper is organized as follows. In Sec.~\ref{sect:theory}
we introduce the formalism that will be used in calculating the
RKKY interaction from the Green's function. In
Sec.~\ref{sect:results} we present our analytic and numeric
results for the coupling strengths of the RKKY interaction in both
undoped and doped monolayer MoS$_2$ sheets. Section~\ref{sect:conclusion}
contains discussions and a brief summary of our main results.

 %topological insulators \cite{ref:TI} which spin is not a conserved quality in them.

\section{Method and Theory}\label{sect:theory}

\begin{figure}
\includegraphics[width=0.5\linewidth]{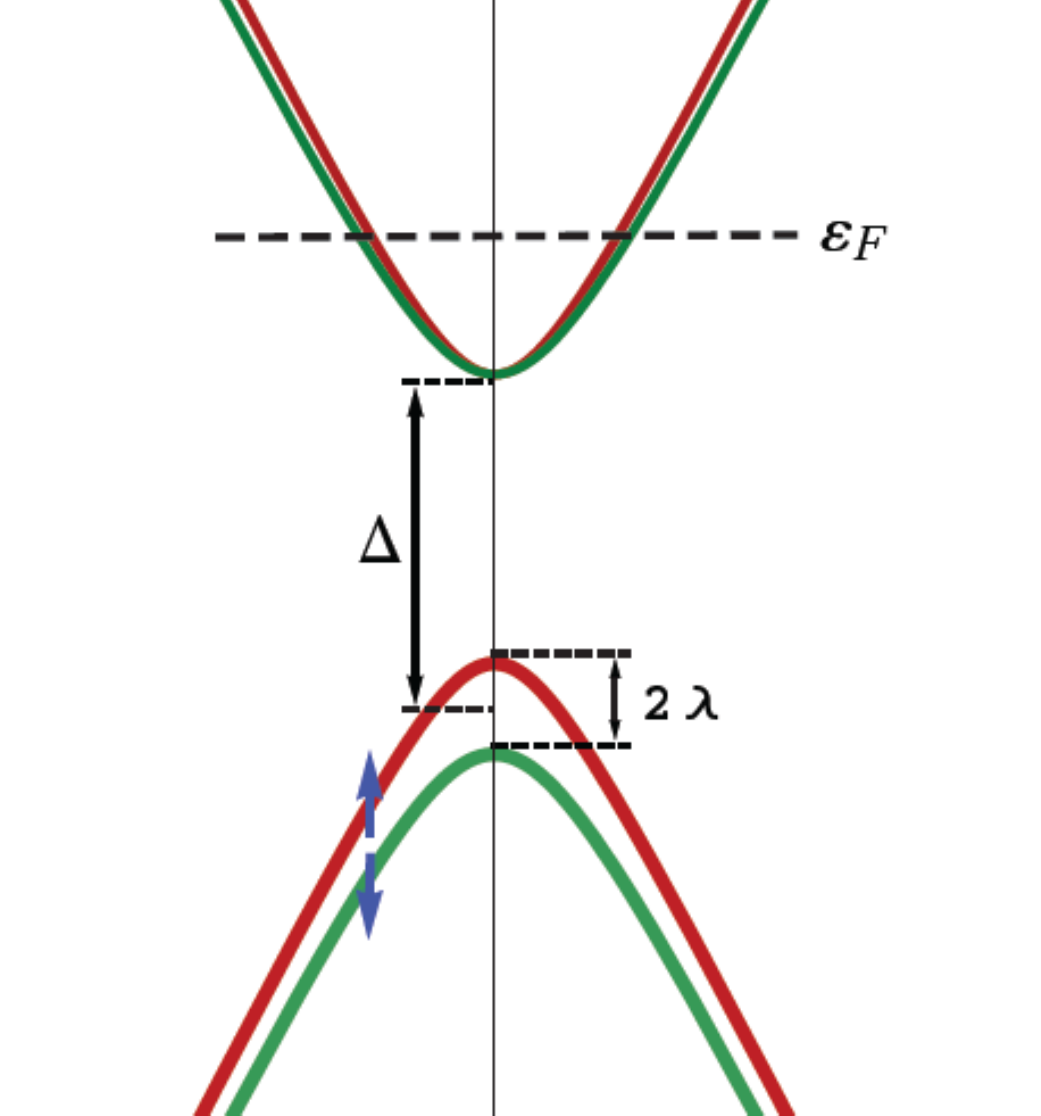}
\caption{(Color online) Schematic picture of the dispersion relation
in momentum space at valley $K$. The valance and conduction bands are separated by a large direct band gap, $\Delta$ and there is a large spin-orbit coupling, $\lambda$, leading to a spin polarization of the valence band, spin up $\uparrow$ and down spin, $\downarrow$. The dispersion relation is given by $E(k)=\frac{\lambda}{2} \tau s_z \pm \sqrt{(atk)^2+(\Delta/2-\lambda \tau s_z /2)^2}$.
\label{fig:schem}}
\end{figure}

The Hamiltonian for a single layer MoS$_2$ including two magnetic impurities can be written as $\mathcal{\hat{H}}=\mathcal{\hat{H}}_0 + \mathcal{\hat{H}}_{int}$, where $\mathcal{\hat{H}}_0$ is the noninteracting Hamiltonian for the single layer MoS$_2$ and $\mathcal{\hat{H}}_{int}$ is the interaction term between impurities and spin of conduction electrons. In the monolayer MoS$_2$, the highest energy valence bands and the lowest energy conduction bands, at the $K$ and $K'$ valleys in momentum space, are mainly of molybdenum $d-$orbital character and spin-orbit interactions split~\cite{xiao} the valence bands by $2\lambda\sim 150$ meV leading to a spin polarization of the valence band. Fig.~\ref{fig:schem} shows the band dispersion of monolayer MoS$_2$ in which spin splitting in conduction band is negligible however a spin splitting in the valance band leads to a spin shift.

The Hamiltonian for MoS$_2$ in a continuum model at $K$ and $K'$ points is written as~\cite{xiao}

\begin{eqnarray}
\mathcal{\hat{H}}^{\tau}_0=at(\tau k_x \hat{\sigma}_x + k_y \hat{\sigma}_y ) +\frac{\Delta}{2} \hat{\sigma}_z - \lambda \tau \frac{ \hat{\sigma}_z-1}{2}\hat{s}_z
\end{eqnarray}
where the index $\tau$ is $\pm 1$ at valley $K$ or $K'$, $\Delta$ is the direct band gap, $\lambda$ is the spin-orbit coupling constant, $a$ is the lattice parameter and $t$ is the hoping integral. $\hat{\sigma}$'s are the Pauli matrices written in pseudospinor $\psi^\dagger=(\Phi^\dagger_c,\Phi^{\tau\dagger}_v)$ where $c$ and $v$ denote the conduction and valence bands, respectively and finally $\hat{s}_z$ is the Pauli matrix for the $z$-component of spin.

Particulary, the noninteracting Hamiltonian in the spinor basis, $\psi^\dagger = (c^\dagger_{k,\uparrow},v^\dagger_{k,\uparrow},c^\dagger_{k,\downarrow},v^\dagger_{k,\downarrow})$ can be written as
\begin{eqnarray}
\mathcal{\hat{H}}^{\tau}_0=\begin{pmatrix}
\Delta/2 & at\tau ke^{-i\tau \theta'} & 0 & 0 \\
at\tau ke^{i\tau \theta'} & -\Delta/2+\lambda \tau & 0 & 0 \\
0 &0 & \Delta/2 & at\tau ke^{-i\tau \theta'} \\
0&0& at\tau  ke^{i\tau \theta'} & -\Delta/2 -\lambda \tau
\end{pmatrix}
\end{eqnarray}
where $\theta'=\tan^{-1}(k_y/k_x)$. The system in question here incorporates two localized magnetic moments whose
interaction is mediated through a large spin-orbit coupled electron gas.
The contact interaction between the spin
of itinerant electrons and two magnetic impurities with magnetic
moments ${\bf I_1}$ and ${\bf I_2}$, located respectively at ${\bf
R}_1$ and ${\bf R}_2$, is given by
\begin{equation}
\mathcal{\hat{H}}_{\rm int} = J_c \sum_{j=1,2} {\bf I}_j\cdot  {\bf s}({\bf R}_j)~,
\end{equation}
where $J_c$ is the coupling constant between conduction
electrons and impurity, ( we set $\hbar=1$ from now on), ${\bf s}({\bf r})=\frac{1}{2} \sum_i
\delta ({\bf r}_i-{\bf r}) \sigmav_i$ is the spin density
operator with ${\bf r}_i$ and $\sigmav_i$ being the
position and vector of spin operators of $i$th electron.

The RKKY interaction which arises from the quantum effects is
obtained by using a second order
perturbation~\cite{ref:RK, ref:Kasuya, ref:Yosida, ref:Imamura} which reads as
\begin{eqnarray} \label{eq:RKKY}
\mathcal{\hat{H}}_{RKKY} = J_c ^2 \sum_{i,j} I_1^i~ \chi_{ij}({\bf R},{\bf R}')~ I_2^j
\end{eqnarray}
where $\chi_{ij}({\bf R},{\bf R}')$ is the spin susceptibility in real space given by
$\chi_{ij}({\bf R},{\bf R}')=\sum_{\alpha\in(c,v)}\chi^{\alpha\alpha}_{ij}({\bf R},{\bf R}')$, in which

\begin{eqnarray}\label{eq:chi1}
\chi^{\alpha\beta}_{ij}({\bf R},{\bf R}')&=&\frac{-1}{2\pi} Tr[\int_{-\infty}^{\varepsilon_F} d\varepsilon \Im m[\sigma_i G^{\alpha\beta}({\bf R},{\bf R}',\varepsilon)\nonumber\\
&\times& \sigma_j G^{\beta\alpha}({\bf R}',{\bf R},\varepsilon)]]
\end{eqnarray}

Here $\alpha$ and $\beta$ denote pseudospin degree of freedom, $\varepsilon_F$ is the Fermi energy, trace is taken over the spin degree of freedom and $\sigma$ is the Pauli spin matrix. $G^{\alpha\beta}({\bf R},{\bf R}';\varepsilon)$ is a $2\times
2$ matrix of the single particle retarded Green's functions in spin
space. In order to calculate the interaction Hamiltonian of
Eq.~(\ref{eq:RKKY}), the form of the electronic single particle
Green's function, $G({\bf R},{\b R}';\varepsilon)=<{\bf
R}|(\varepsilon+i0^{+}-\mathcal{\hat{H}}_0)^{-1}|{\bf R}'>$
is needed. To calculate the retarded Green's function in real
space, its Fourier components in momentum space might be first
obtained.
\begin{widetext}
\begin{equation}
G(k,\varepsilon)=\begin{pmatrix}
\frac{\varepsilon+\Delta/2-\lambda \tau}{(\varepsilon-\Delta/2)(\varepsilon+\Delta /2-\lambda \tau)-a^2t^2k^2} & \frac{at\tau ke^{-i\tau\theta'}}{(\varepsilon-\Delta/2)(\varepsilon+\Delta /2-\lambda \tau)-a^2t^2k^2} &0&0\\
 \frac{at\tau ke^{i\tau\theta'}}{(\varepsilon-\Delta/2)(\varepsilon+\Delta /2-\lambda \tau)-a^2t^2k^2} & \frac{\varepsilon-\Delta/2}{(\varepsilon-\Delta/2)(\varepsilon+\Delta /2-\lambda \tau)-a^2t^2k^2}&0&0\\
 0&0 &  \frac{\varepsilon+\Delta/2+\lambda \tau}{(\varepsilon-\Delta/2)(\varepsilon+\Delta /2+\lambda \tau)-a^2t^2k^2} & \frac{at\tau ke^{-i\tau \theta'}}{(\varepsilon-\Delta/2)(\varepsilon+\Delta /2+\lambda \tau)-a^2t^2k^2} \\
 0&0 & \frac{at\tau k e^{i\tau \theta'}}{(\varepsilon-\Delta/2)(\varepsilon+\Delta /2+\lambda \tau)-a^2t^2k^2} &
 \frac{\varepsilon-\Delta/2}{(\varepsilon-\Delta/2)(\varepsilon+\Delta /2+\lambda \tau)-a^2t^2k^2}
\end{pmatrix}
\end{equation}
\end{widetext}
Generally, the energy dispersion of the noninteracting system can be obtained by the pole of the Green's function and it reads as
$E(k)=\frac{\lambda}{2} \tau s_z \pm \sqrt{(atk)^2+(\Delta/2-\lambda \tau s_z /2)^2}$.
The retarded Green's function are diagonal in the spin space, $G_{cc}(q,\varepsilon)=
\begin{pmatrix}
\frac{\alpha_-}{\beta_- - q^2} &0\\
0 & \frac{\alpha_+}{\beta_+ - q^2}
\end{pmatrix} $ and
$G_{vv}(q,\varepsilon)=
\begin{pmatrix}
\frac{\alpha}{\beta_{-} - q^2} &0\\
0 & \frac{\alpha}{\beta_{+} - q^2}
\end{pmatrix}$ where $\alpha_\pm = (\varepsilon+\Delta/2\pm \lambda)/(at)^2$, $\alpha = (\varepsilon-\Delta/2)/(at)^2$ and $\beta_\pm = (\varepsilon-\Delta/2)(\varepsilon+\Delta/2\pm \lambda)/(at)^2$. Taking the Fourier transform of the matrices, we thus find the real space Green's function matrix
\begin{eqnarray}
G_{cc}({\bf R},0,\varepsilon)=\frac{1}{(2\pi)^2}&&\int d^2 q [e^{i{\bf K}\cdot{\bf R}}e^{i{\bf q}\cdot{\bf R}}G_{cc}(q(K)\varepsilon)+\nonumber\\
&&e^{i{\bf K}'.{\bf R}}e^{i{\bf q}.{\bf R}}G_{cc}(q(K'),\varepsilon)]
\end{eqnarray}
where $q(K)$ or $q(K')$ is the wavevector near the Dirac point $K$ or $K'$, respectively. The integral can be simply calculated and it thus reads as
\begin{eqnarray}\label{eq:gc}
&&G_{cc}({\bf R},0,\varepsilon)=\nonumber\\
&&\begin{pmatrix}
e^{i{\bf K}\cdot{\bf R}}g_{c-}+e^{i{\bf K}'\cdot{\bf R}}g_{c+} &0\\
 0& e^{i{\bf K}\cdot{\bf R}}g_{c+}+e^{i{\bf K}'\cdot{\bf R}}g_{c-}\\
 \end{pmatrix}
\end{eqnarray}
where $g_{c\pm}(|{\bf R}|)=-\alpha_{\pm}K_0(\sqrt{-\beta_\pm}|{\bf R}|)/(2\pi)$ in which $K_0(x)$ is the modified Bessel function of the second kind. We can follow conveniently the same procedure discussed above to obtain $G^0_{cc}(0,{\bf R},\varepsilon)=G^0_{cc}(-{\bf R},0,\varepsilon)$.
Moreover, the retarded Green's function $G^0_{vv}({\bf R},0,\varepsilon)$ is obtained the same as the channel $(cc)$ with replacing $g_{v\pm}(|{\bf R}|)=-\alpha K_0(\sqrt{-\beta_{\pm}}|{\bf R}|)/(2 \pi)$ instead of $g_{c\pm}(|{\bf R}|)$.

By inserting the retarded Green's functions in Eq.~(\ref{eq:RKKY}), and
taking the trace over spin degree of matrices, the RKKY
Hamiltonian simplifies to

\begin{eqnarray}
\mathcal{\hat{H}}_{RKKY} &=& J_c^2\chi_{xx}(I_{1x}I_{2x}+I_{1y}I_{2y}) \nonumber\\
&+&J_c^2 [\chi_{xy}(I_1\times I_2)_z+ \chi_{zz}I_{1z}I_{2z}]
\end{eqnarray}
which indicates that the RKKY interaction consists of three different terms: the Heisenberg like $I_1 \cdot I_2$, Ising $I_{1z} I_{2z}$ and Dzyaloshinsky-Moriya $(I_1 \times I_2)_z$ terms. Accordingly, the RKKY
Hamiltonian can be written as
\begin{eqnarray}\label{eq:eta}
\mathcal{\hat{H}}_{RKKY}=J_H {\hat H}_H + J_I {\hat H}_I + J_{DM} {\hat H}_{DM}
\end{eqnarray}
where $J_i=-\frac{J_c^2}{2\pi}\int_{-\infty}^{\varepsilon_F} d\varepsilon \Im m~\eta_i$ and
\begin{eqnarray}
\eta _H&=&\sum_{a\in(c,v)} 4g_{a+}g_{a-}+2(g_{a+}^2+g_{a-}^2)\cos(({\bf K}-{\bf K}')\cdot{\bf R})\nonumber\\
\eta _I&=&\sum_{a\in(c,v)} 2(g_{a+}-g_{a-})^2(1-\cos(({\bf K}-{\bf K}')\cdot{\bf R}))\nonumber\\
\eta _{DM}&=&\sum_{a\in(c,v)} -2(g_{a+}^2-g_{a-}^2)\sin(({\bf K}-{\bf K}')\cdot{\bf R})
\end{eqnarray}
The resulting RKKY interaction consists of three quite different interactions. The Heisenberg and Ising interactions favor a collinear alignment of localized spins. On the contrary, the DM coupling favors a noncollinear alignment of localized spins.
The same form of Hamiltonian has been obtained in 2D electron gas in the presence of Rashba spin-orbit coupling~\cite{ref:Imamura} and in the topological insulators, on which
magnetic impurities exhibit a frustrated RKKY interaction with two
possible phases of ordered ferromagnetic phase and a disordered
spin glass phase~\cite{ref:TI1}. Importantly, the existence of $J_{DM}$ in those systems is a consequence of spin changing term however in the monolayer MoS$_2$ system, the DM coupling is a consequence of the spin-orbit coupling and the existence of two inequivalent valleys separated by $({\bf K}-{\bf K}')$. Consequently, the DM term results in noncollinearity of two spinors interacting to each other.
In addition, all coupling terms in the Hamiltonian for the monolayer MoS$_2$ system depend on the direction between magnetic impurities.

\section{Numerical results }\label{sect:results}

In this section, we present in the following our main results for the numerical evaluation of the RKKY exchange
coupling in massive Dirac fermions
by analyzing the above calculated integrals of
$J_H$, $J_{DM}$ and $J_I$. We set~\cite{xiao} $a=3.193${\AA}, $\Delta=1.66$eV, $2\lambda=150$meV and $t=1.1$eV
in all our results presented in this Section. we also consider two valleys at $K= (2\pi)(1,\sqrt{3})/(3\sqrt{3}a) $ and $K'= (2\pi)(-1,\sqrt{3})/(3\sqrt{3}a) $.

\begin{figure}
\includegraphics[width=1.1\linewidth]{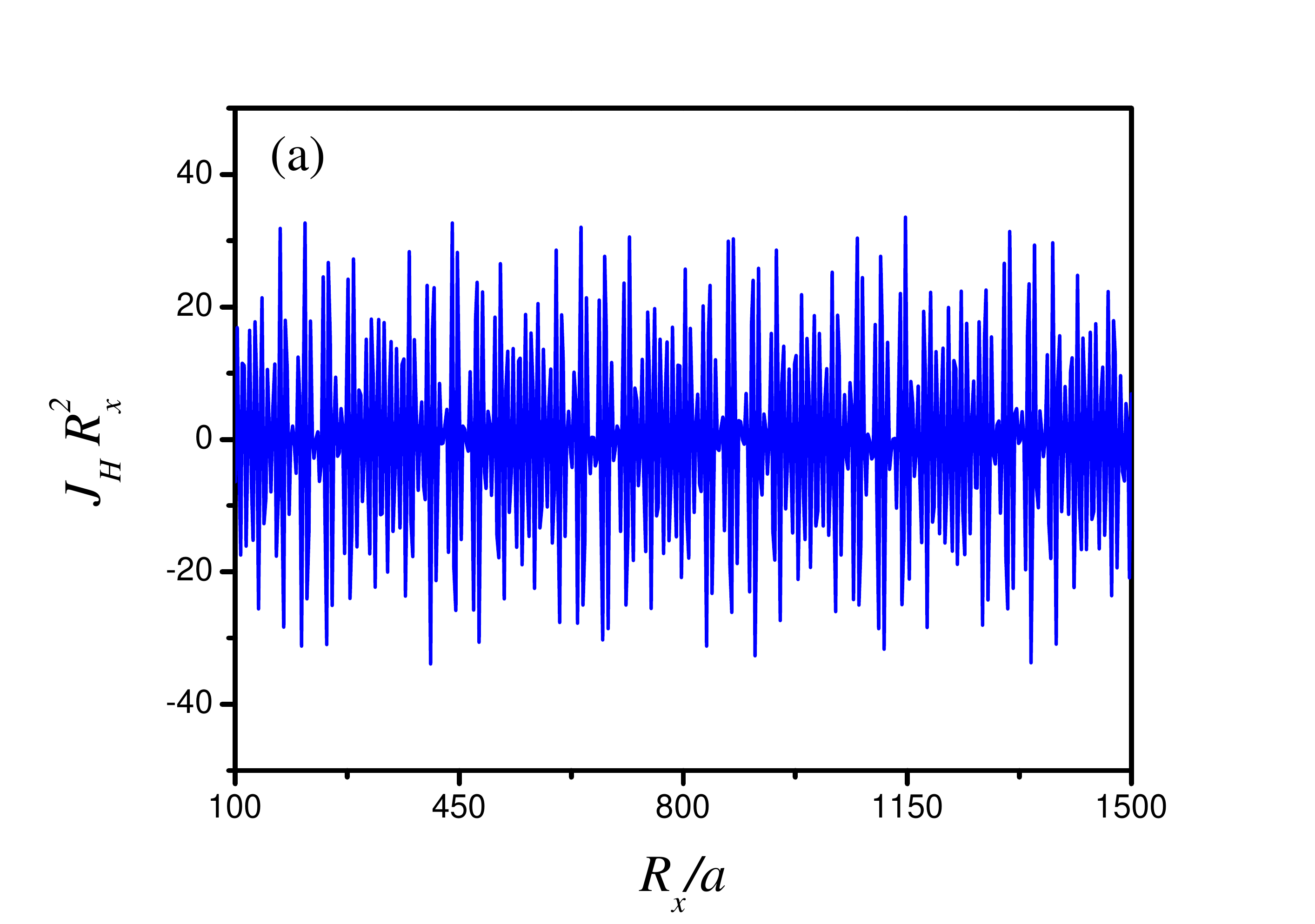}
\includegraphics[width=1.1\linewidth]{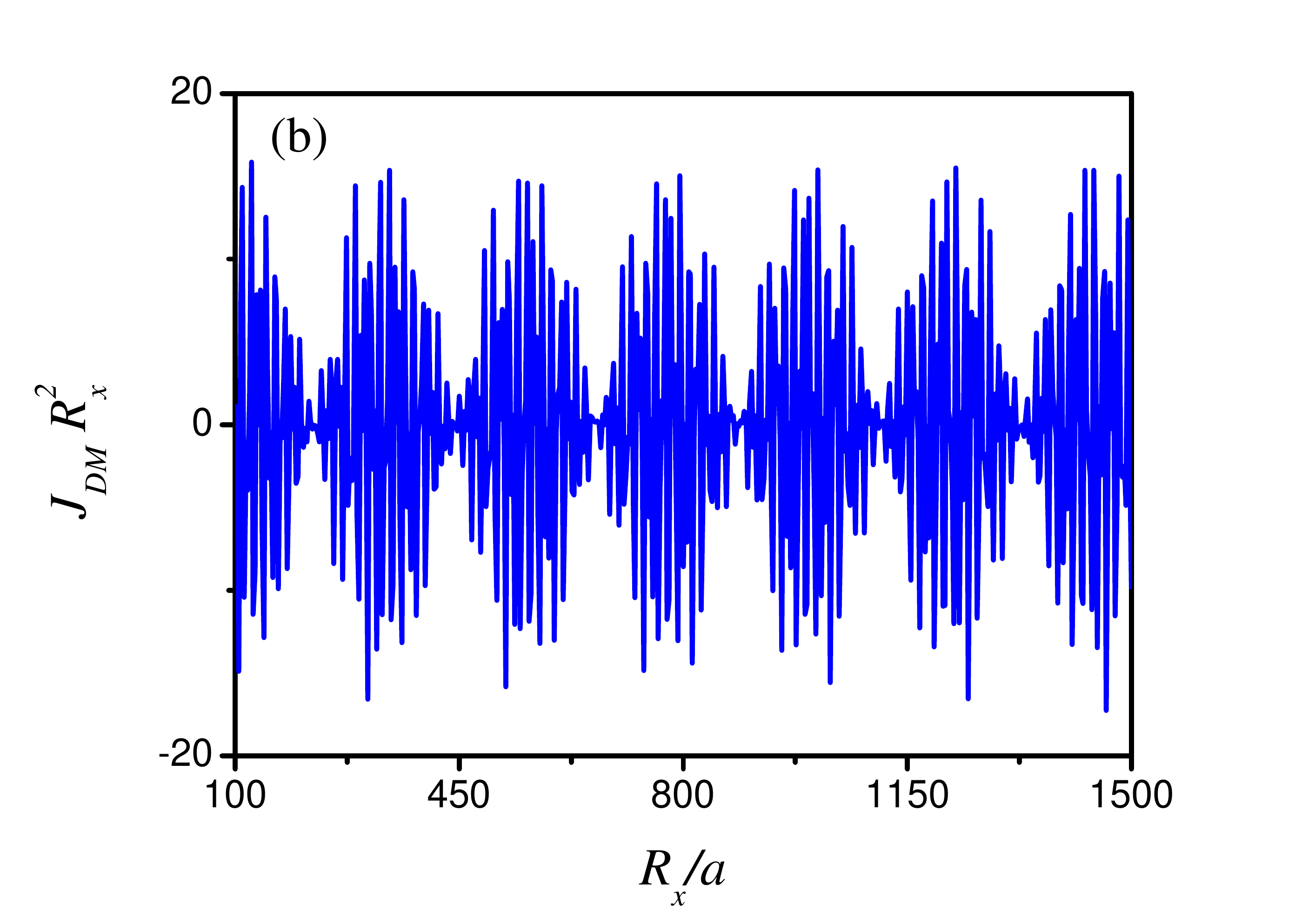}
\includegraphics[width=1.1\linewidth]{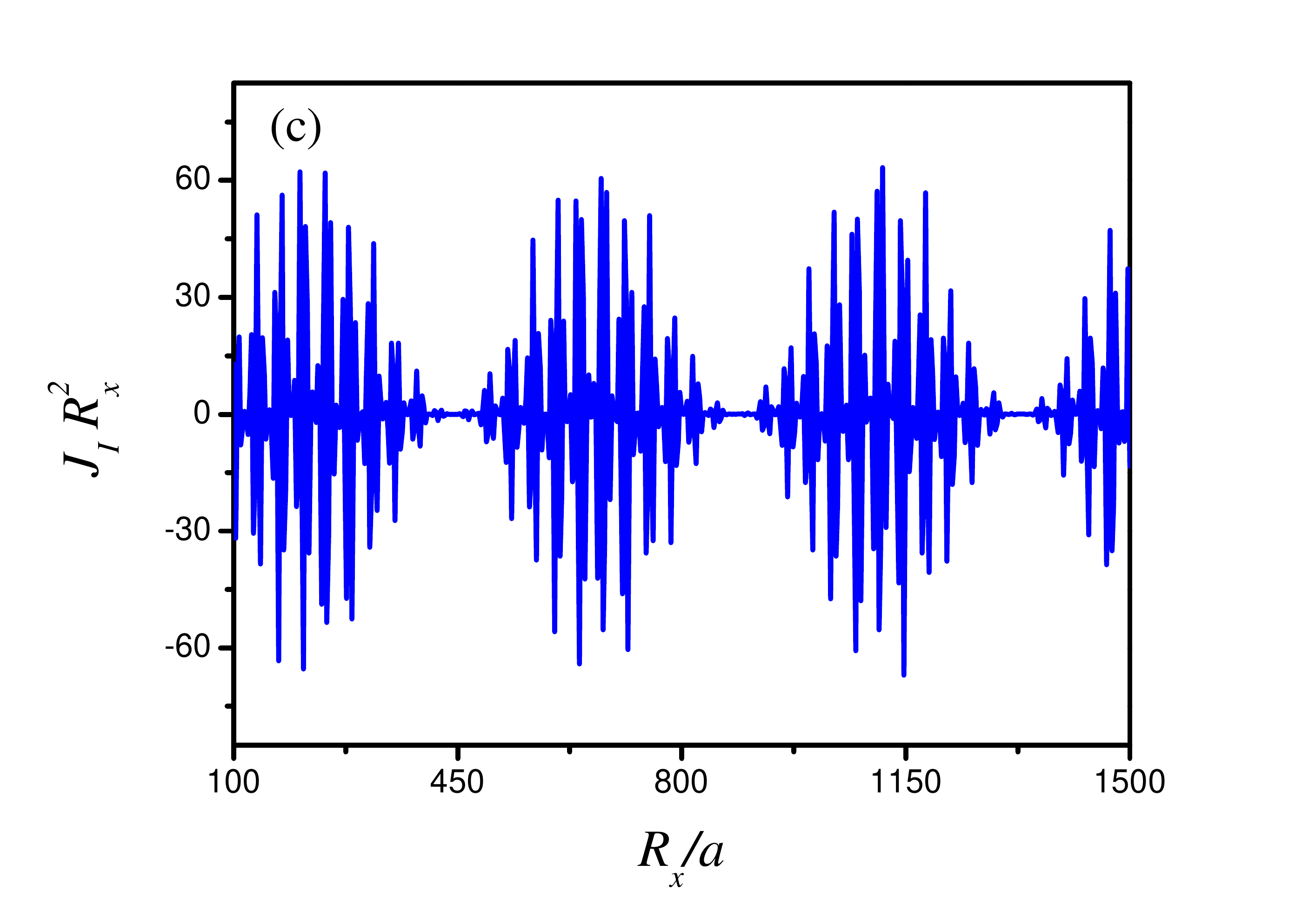}
\caption{(Color online) Long range behavior of RKKY interaction terms, $J_i$ for $i=H, DM$ and $I$, times to $R^2_x$ scaled by $J^2_c/(2\pi)^3 a^2$ as a function of the impurities distance along $x$ direction for the Fermi energy which intersects the conduction band
at $\varepsilon_{\rm F}=\Delta/2+\lambda$. The behavior of all $J_i$ interactions are
determined by a superposition of two modified Bessel functions (see the text) with two
different periods determined by arguments in $\beta_\pm$. Moreover the RKKY interaction terms in doped MoS$_2$ decay as $R^{-2}$ at long-range distance.
\label{fig:longrange}}
\end{figure}

At finite value of the Fermi energy, electron or hole doped,
more complicated behavior of the RKKY coupling can
be occurred. In this case, the behavior of all $J_i$ interactions are
determined by a superposition of two modified Bessel functions with two
different periods determined by arguments in $\beta_\pm$.
As the result, we observe that for doped MoS$_2$, oscillations of
$J_i$ exhibit a beating pattern. Fig.~\ref{fig:longrange}(a) shows this
beating behavior of integral, $J_{H}$ scaled by $J^2_c/(2\pi)^3 a^2$ as a function of the
impurities distance along $x$ direction ( zigzag direction) for the Fermi energy which intersects the conduction band
at $\varepsilon_{\rm F}=\Delta/2+\lambda$.
Fig.~\ref{fig:longrange}(b) and (c) show the similar behavior for integral
$J_{DM}$ and $J_I$, respectively as a distance along $x$ direction. Our numerical results
show that all coupling interactions decay as $R^{-2}$ at finite Fermi energy.
As it is obvious from Eq.~(\ref{eq:eta}), in a long-range regime, $J_I>J_H>J_{DM}$ and moreover the coupling $J_{DM}$ and $J_{I}$ vanish when $\lambda \rightarrow 0$.
Therefore, the existence of DM and Ising interactions is a consequence of spin-orbit coupling without having Rashba or Dresselhaus interactions. Besides, since $g_{c\pm}$ are coupled to the Dirac points differently as diagonal elements in Eq.~(\ref{eq:gc}), the DM coupling is nonzero as long as one has the spin-orbit coupling together with two inequivalent valleys.
It should be noticed that there are some locations along $x$ direction for which $J_{DM}\gg J_{H}$, since they have two different periods and this leads to a strong rotation even at long distance.

\begin{figure}
\includegraphics[width=1.1\linewidth]{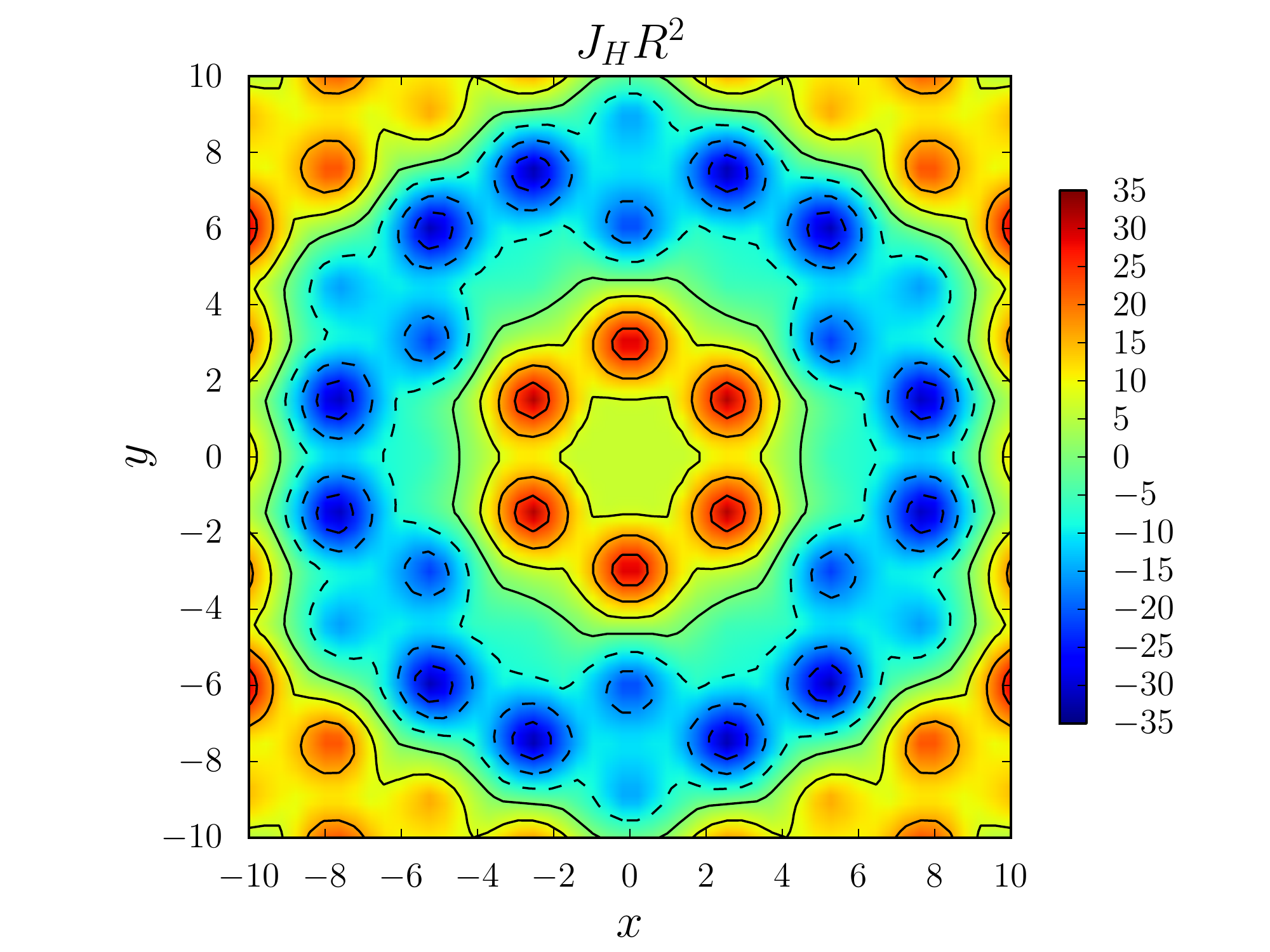}
\includegraphics[width=1.1\linewidth]{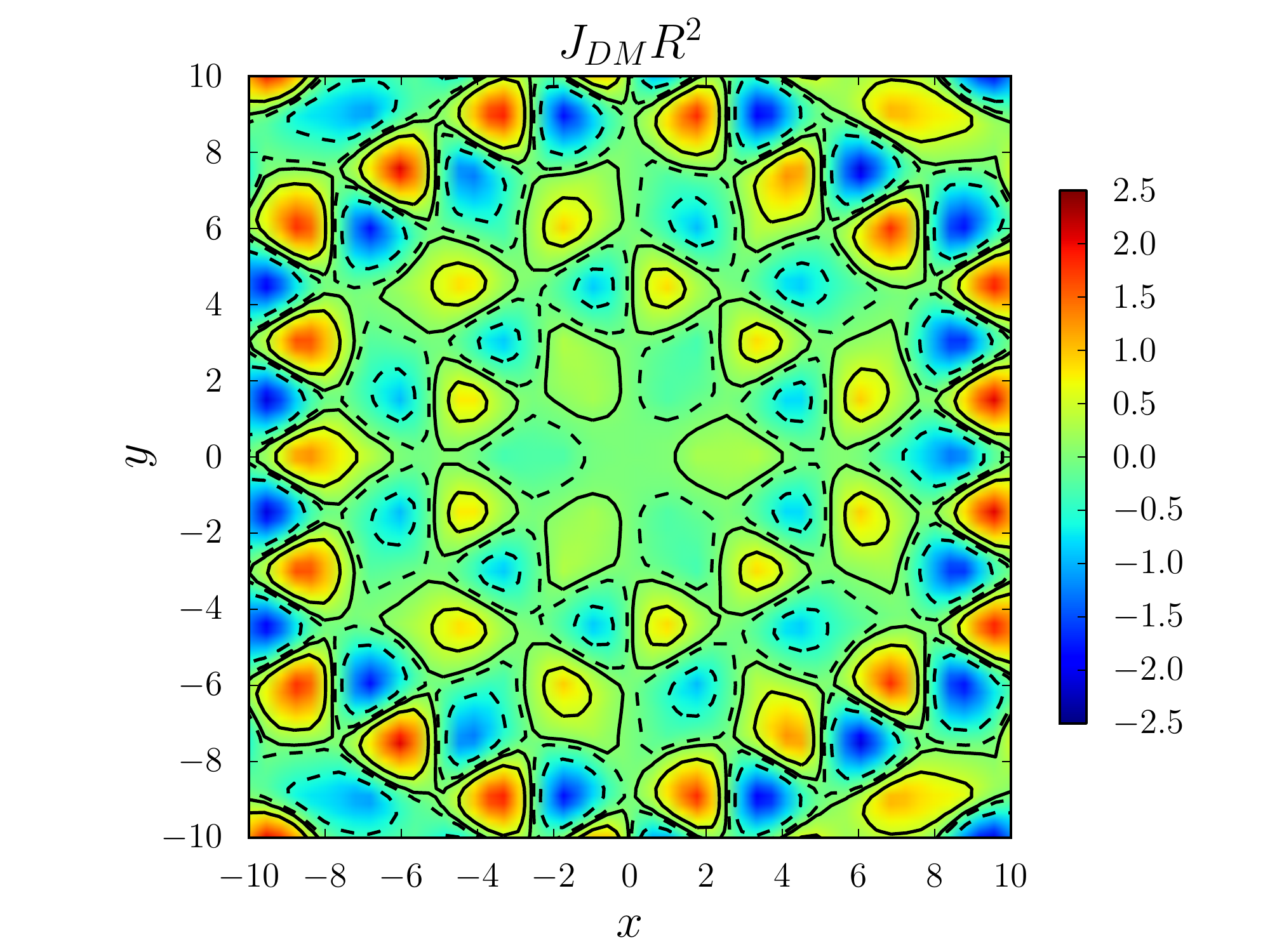}
\includegraphics[width=1.1\linewidth]{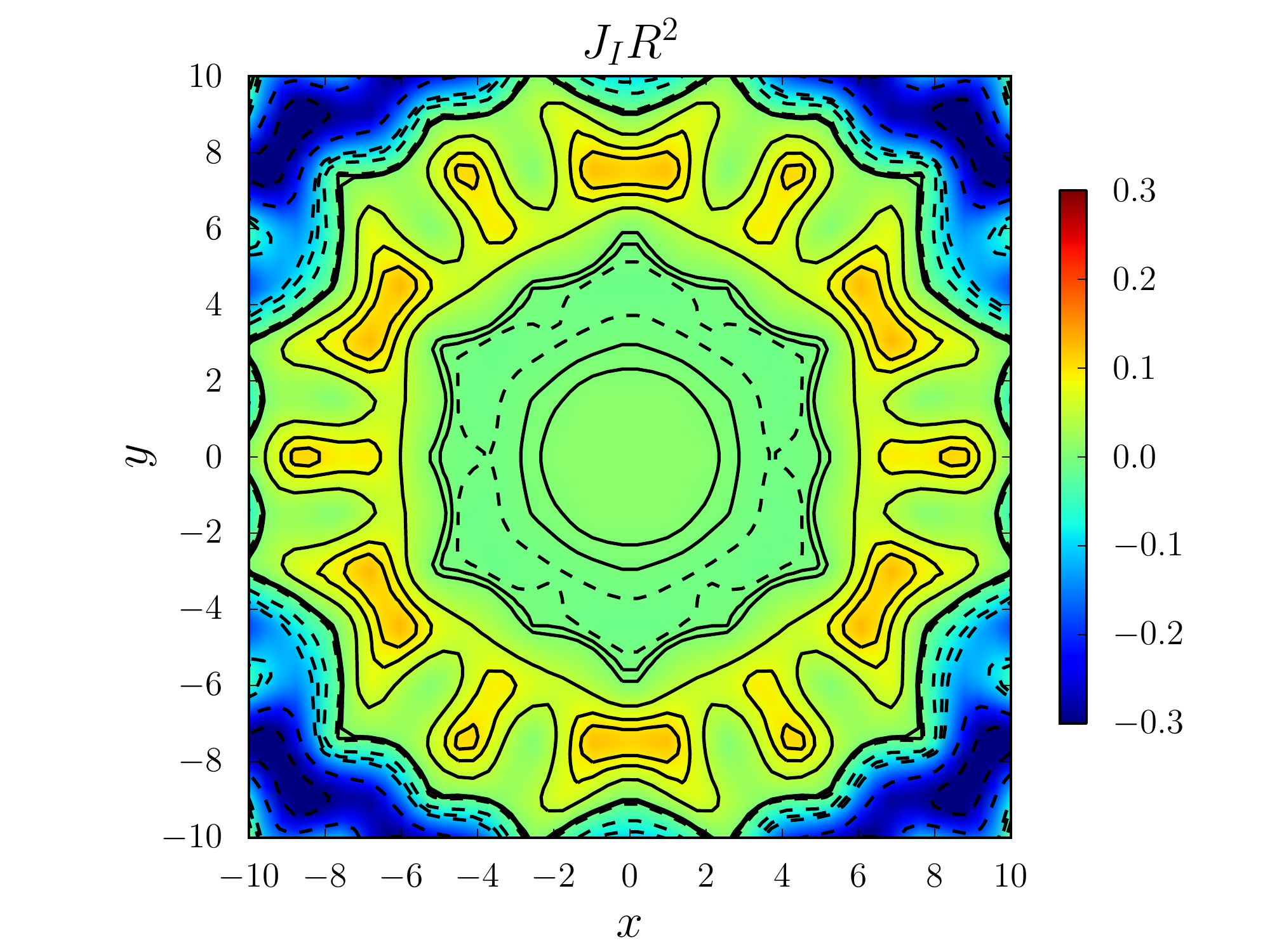}
\caption{(Color online) Short range contour plot of RKKY interaction terms, $J_i$ for $i=H, DM$ or $I$ scaled by $J^2_c/(2\pi)^3 a^2$ as a function of the impurities distance in $(x,y)$ plane scaled by $a$ for the Fermi energy $\varepsilon_{\rm F}=\Delta/2+\lambda$. One of the spins is fixed at origin and another can be located on lattice points. Solid lines denote boundaries with positive values of the interaction term while dashed lines refer to its negative values. The results show clearly the three-fold symmetric.
\label{fig:contour}}
\end{figure}

Notice that in the MoS$_2$ system, the RKKY interaction is direction dependant which is in contrast to the results obtained in the topological insulators or 2D electron gas systems. Since the RKKY interaction is a weakly interaction and decays as $R^{-2}$ in the MoS$_2$ system, it would be good to look at its short-range behavior. Fig.~\ref{fig:contour}, shows the angel dependance of $J_i$ in contour plot scheme for a short distance and for $\varepsilon_{\rm F}=\Delta/2+\lambda$. Here, the system is electron doped and we assume that one of the spins is fixed at origin and another can be located on lattice points. Close to the impurity positions, the effective interaction, comes mainly from $J_H$ interaction, is ferromagnetic and becomes antiferromagnetic at a distance comparable to $\sim 2a$. As expected, the amplitude of oscillations decrease with increasing distance between the local moment impurities. Moreover, the symmetries of the band structure which is a three-fold symmetric reflects in the susceptibility and thus in the magnetic interactions. The three-fold symmetric is clear in this figure.

\begin{figure}
\includegraphics[width=1.1\linewidth]{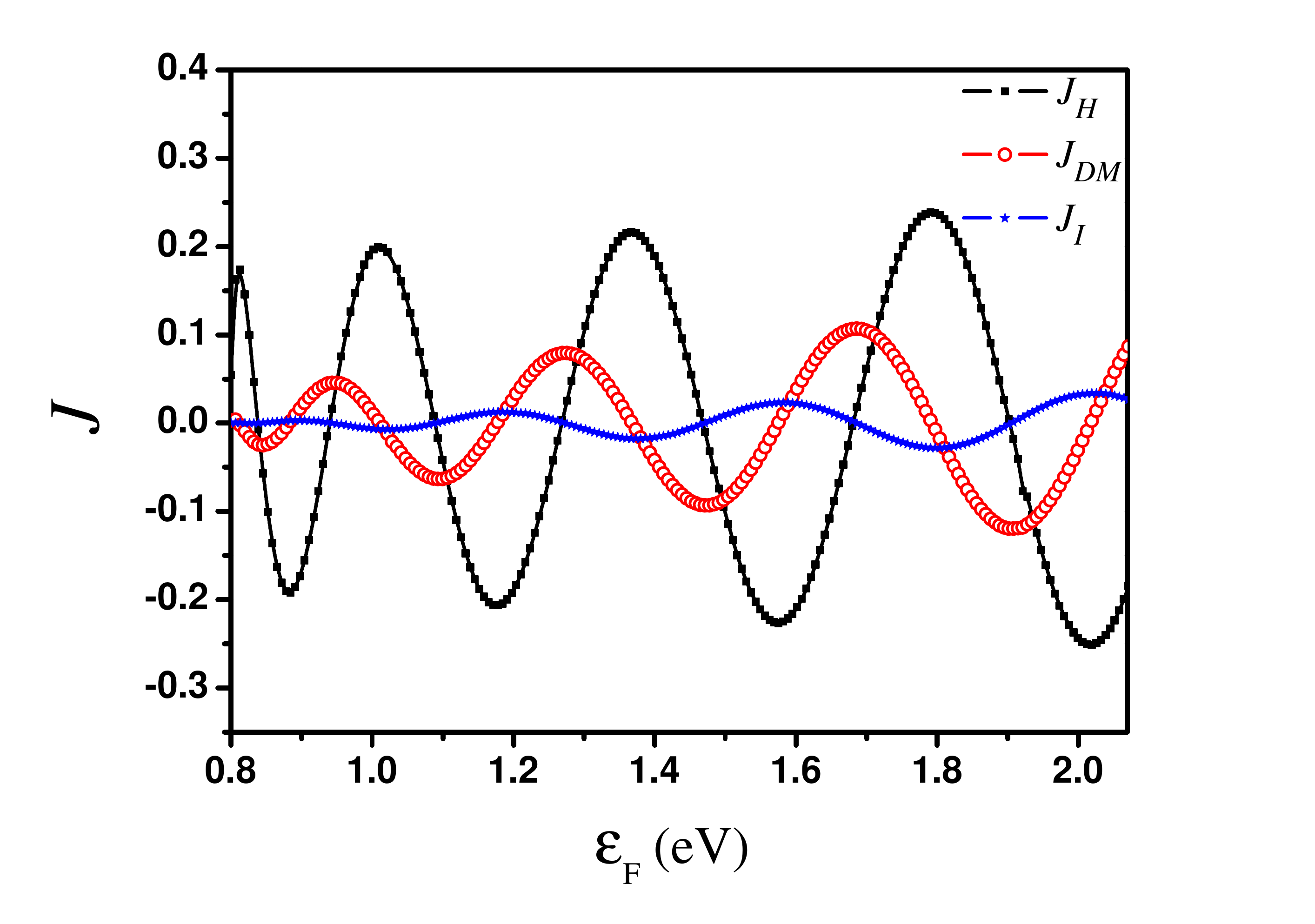}
\caption{ (Color online) RKKY interaction terms, $J_i$ for $i=H, DM$ or $I$ scaled by $J^2_c/(2\pi)^3 a^2$ as a function of electron energy doping in units of eV at ${\bf R}=(4\sqrt{3}a, 0)$. Since there is a phase difference between $J_i$'s, for certain value of the Fermi energies, there are situations in which $J_{DM}> J_{H}$. 
\label{fig:condef}}
\end{figure}

For a situation where the Fermi energy has an intersection to the valance bands (hole doped case), the RKKY interaction terms exhibiting oscillatory behavior structures however, the magnitude of the interactions are different from those results obtained for the case of the electron dope system.
An intriguing manner of the RKKY interaction is its dependence of the Fermi energy values. The Fermi energy dependence of the RKKY interaction is shown in Fig.~\ref{fig:condef} for which the Fermi energy always intersects the conduction band and for a certain value of distance between two impurities, ${\bf R}=(4\sqrt{3}a, 0)$. The interaction amplitudes increase by increasing the Fermi energy and the period of the oscillations remain constant. It is expected that the values of $J_{DM}$ and $J_I$, in the short-range, be small since the contribution of the spin splitting in the conduction band is negligible. Owing to a phase difference between $J_i$'s, for certain values of the Fermi energy, there are situations in which $J_{DM}> J_{H}$. Interestingly enough, for a certain values of the Fermi energy and the distance between two magnetic impurities, $J_{DM}> J_{H}$ and leads to a strong rotation between spins of impurities. Accordingly, by designing an arbitrary artificial lattice with magnetic impurities (e.g. triangular or square with arbitrary lattice constant), by properly adjusting the distance ${\bf R}$ using the scanning tunneling microscope techniques, we will be able to control each coupling term in the RKKY interaction and therefore the monolayer MoS$_2$ would be a quite good platform to exhibit different spin lattices and interactions.

\begin{figure}
\includegraphics[width=1.1\linewidth]{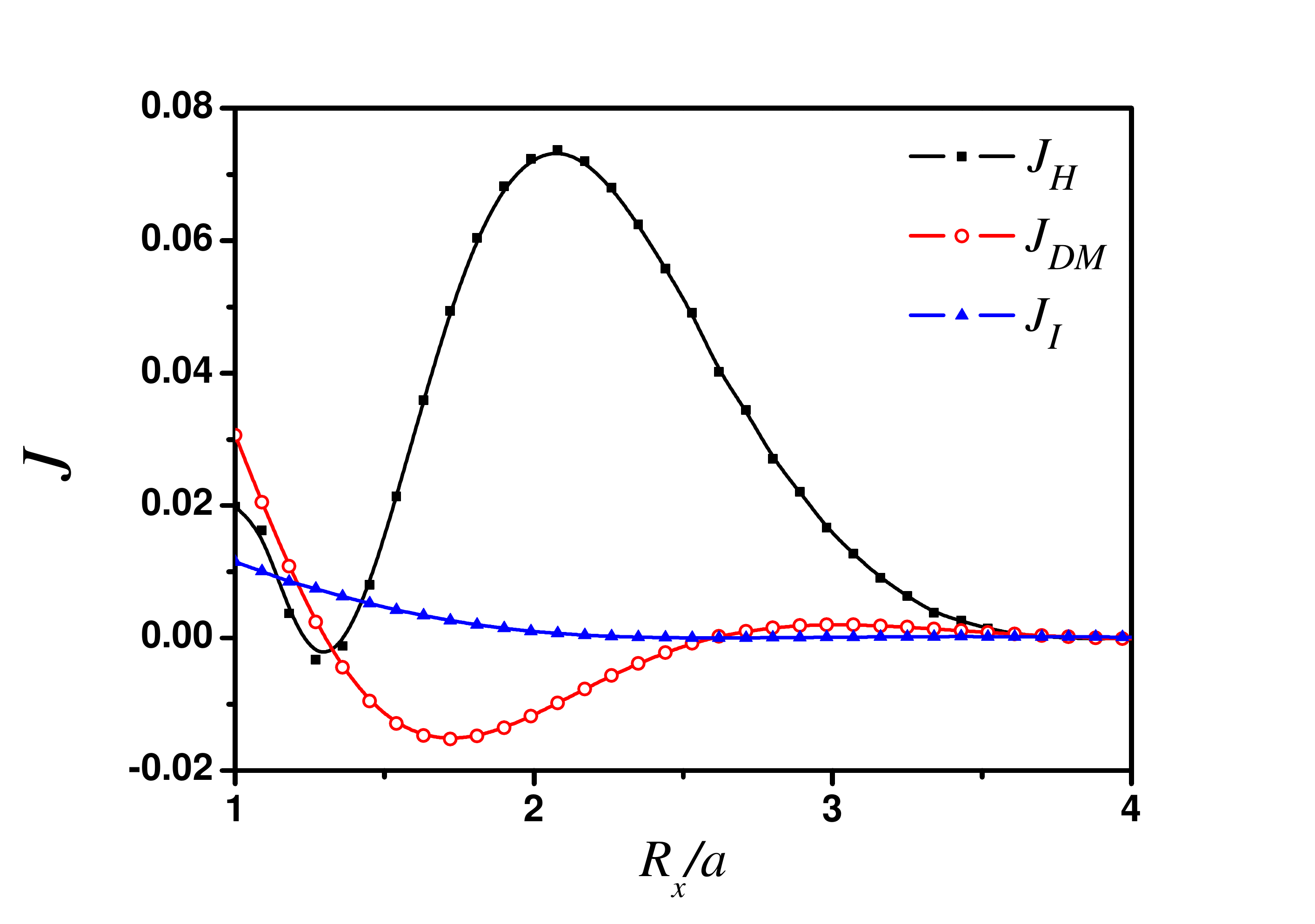}
\caption{ (Color online) Short-range behavior of RKKY interaction terms, $J_i$ for $i=H, DM$ or $I$ scaled by $J^2_c/(2\pi)^3 a^2$ as a function of the impurities distance along $x$ direction for undoped monolayer MoS$_2$ system where $\varepsilon_{\rm F}=0$. $J_i$ coupling interactions tend to zero quickly and moreover, $J_{H}$ and $J_I$ are mostly negative and the interaction is an antiferromagnetic.
\label{fig:gapR}}
\end{figure}

On the other hand, for an undoped MoS$_2$ system, the response of electron decreases exponentially and therefore the RKKY interaction terms fall off rapidly. This phenomena is the same behavior which occurs in the topological insulators when the Fermi energy lies in a gap~\cite{ref:TI}. Fig.~\ref{fig:gapR} shows the RKKY interaction terms for an undoped monolayer MoS$_2$. Those interactions decay fast which denote that the
interactions are rather short-ranged~\cite{ref:BR} and tend to zero quickly. Moreover, $J_{H}$ and $J_I$ are mostly positive which shows that the coupling between the moments remains antiferromagnetic-like for all $Rs$.

\begin{figure}
\includegraphics[width=1.1\linewidth]{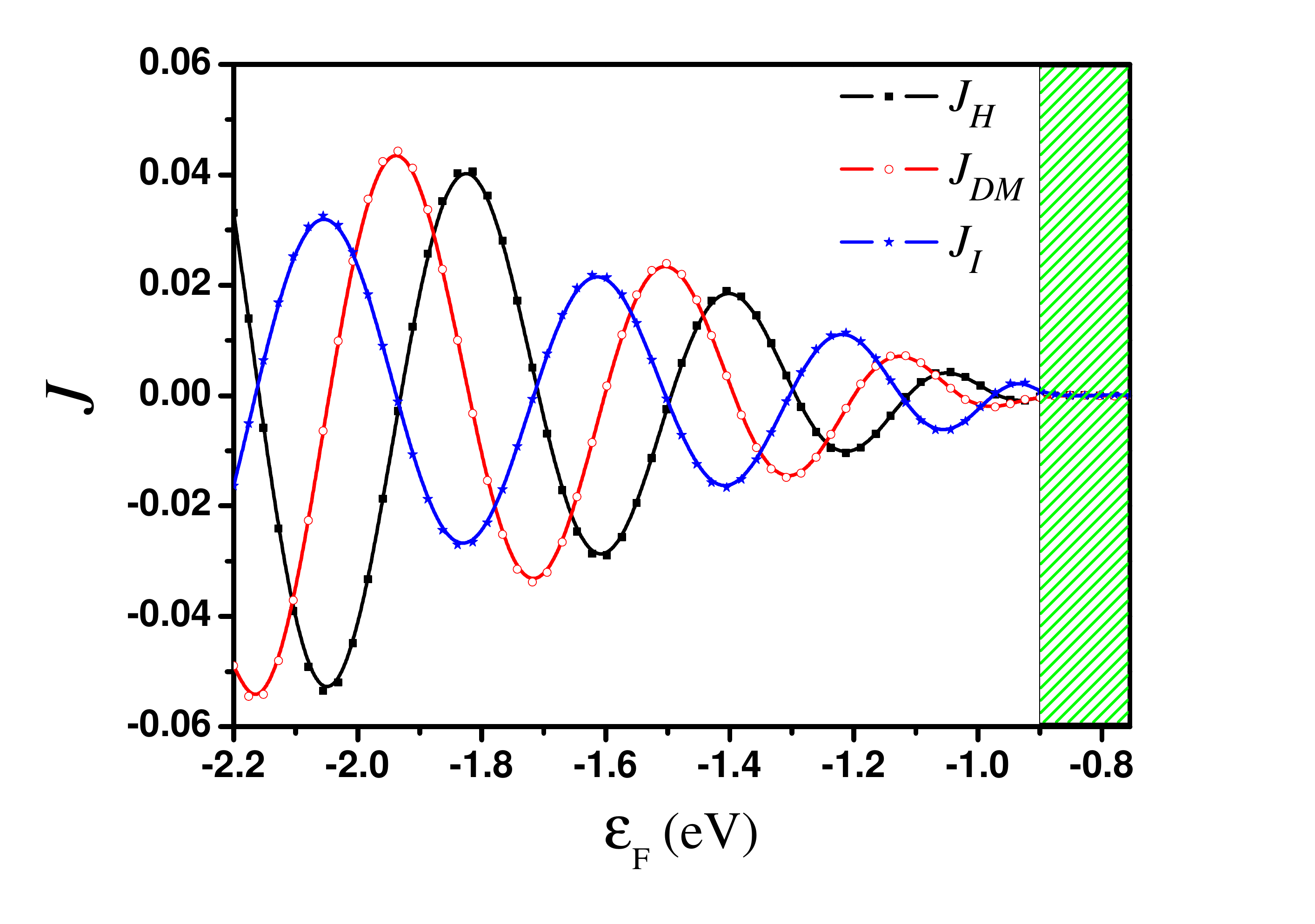}
\caption{ (Color online) RKKY interaction terms, $J_i$ for $i=H, DM$ or $I$ scaled by $J^2_c/(2\pi)^3 a^2$ as a function of the hole energy dope where $\varepsilon_{\rm F}<-\Delta/2-\lambda$ for ${\bf R}=(4\sqrt{3}a, 0)$. The RKKY interaction terms oscillate and their amplitudes suppress by increasing the Fermi energy till it reaches to the top valance band illustrated by a shadowed layer.
\label{fig:vallanceef}}
\end{figure}

In the monolayer MoS$_2$, there are two valance bands which are separated out by the spin-orbit interaction and furthermore, each band has a different $s_z$ value ( see Fig.~1). The top valance band starts at $\varepsilon = -\Delta/2+\lambda$ whereas the another band starts at $\varepsilon = -\Delta/2-\lambda$. Fig.~\ref{fig:vallanceef} shows the behavior of the RKKY interaction terms with respect to the Fermi energy in the region where the valance band is occupied. For the case that $\varepsilon_{\rm F}<-\Delta/2-\lambda $, an important result will be occurred. First of all, the RKKY interaction terms exhibit oscillatory structures and their amplitudes suppress by increasing the Fermi energy. This behavior occurs till the Fermi energy reaches to the top valance band illustrated by a shadowed layer in Fig.~\ref{fig:vallanceef} and afterwards the RKKY interaction terms decay rapidly. The reason for such decaying comes from the fact that the one spin channel, for the such Fermi energy value, is blocked and therefore, the magnetic spin can not be coupled to the host electrons. In other words, in this case, the spin susceptibility of the system tends to zero since the spin relaxation time of the monolayer MoS$_2$ is quite long~\cite{xiao} when the intervalley scattering is ignored.

\subsection{Twisted angle: A classical picture}

As discussed before, the DM term causes a twisted interaction between two magnetic adatoms. In order to calculate the equilibrium angel of the interaction, we simplify the RKKY interaction Hamiltonian in the spherical coordinate as:
\begin{eqnarray}\label{eq:h1}
\mathcal{\hat{H}}_{RKKY}=I_1 I_2(J_H+J_I) \cos\theta_1\cos\theta_2\nonumber\\
+I_1 I_2\sin\theta_1\sin\theta_2(J_H\cos\phi+J_{DM}\sin\phi)
\end{eqnarray}
where $\theta_{1(2)}$ denotes the angle between spin along $z$ direction and $\phi=\phi_2-\phi_1$. The equilibrium conditions, $\partial \mathcal{\hat{H}}_{RKKY}/\partial \theta_1=0$,$\partial \mathcal{\hat{H}}_{RKKY}/\partial \theta_2=0$ and $\partial \mathcal{\hat{H}}_{RKKY}/\partial \phi=0$, lead to the following equations:
\begin{widetext}
\begin{eqnarray}\label{eq:h2}
-(J_H+J_I)\cos\theta_2\sin\theta_1 +\sin\theta_2\cos\theta_1(J_H\cos\phi+J_{DM}\sin\phi)&=&0 \nonumber \\
-(J_H+J_I)\cos\theta_1\sin\theta_2 +\sin\theta_1\cos\theta_2(J_H\cos\phi+J_{DM}\sin\phi)&=&0 \nonumber \\
\sin\theta_1\sin\theta_2(-J_H\sin\phi+J_{DM}\cos\phi)&=&0 \nonumber\\
\end{eqnarray}
\end{widetext}
There are solutions for Eq.~(\ref{eq:h2}) at the boundary of the square $(\theta_1-\theta_2)$ plane which are conveniently given by $\pi( 1,0)$, $\pi(0, 1)$, $\pi(0, 0)$ and $\pi(1, 1)$. For those solutions at the boundaries, possible configurations of the two magnetic impurities are only ferromagnetic- or antiferromagnetic-like configurations. Moreover, there is just one extremum point at $\theta_1=\theta_2=\theta=\pi/2$ between $\theta=0$ and $\pi$. The states of the system are described by a maximum, minimum or saddle behaviors of the extremum. An intriguing case can be occurred for the situation that the extremum is a local minimum. In this case, the equilibrium condition leads to the case that the moments lie in the sample's plane with an angle $\phi=\tan^{-1}(J_{DM}/J_H)$ between two magnetic moments.

In order to describe the extremum behavior at point $\theta=\pi/2$, we do need to evaluate the positive definite condition of the Hessian matrix of Eq.~(\ref{eq:h1}) throughout a neighbor of $\theta=\pi/2$, which is the second derivative of the RKKY Hamiltonian with respect to free variables~\cite{book}, and it yields as
\begin{eqnarray}
D_1&=&-sign(J_H)\sqrt{J_H^2+J_{DM}^2} \nonumber \\
D_2&=&J_{DM}^2-J_I(2J_H+J_I) \nonumber
\end{eqnarray}
For a case that $D_1>0$ and $D_2>0$, the critical point $\theta=\pi/2$ is a local minimum of the system of which we would like to explore the system. However, for the situation in which $D_1<0$ and $D_2>0$ the critical point is a local maximum of the system. A saddle point, on the other hand, is described by condition that $D_1>0$ and $D_2<0$ or $D_1<0$ and $D_2<0$. For two latest cases, the equilibrium state of the system is obeyed by the boundary solutions which given by ferromagnetic- or antiferromagnetic-like configurations.

\begin{figure}
\includegraphics[width=1.1\linewidth]{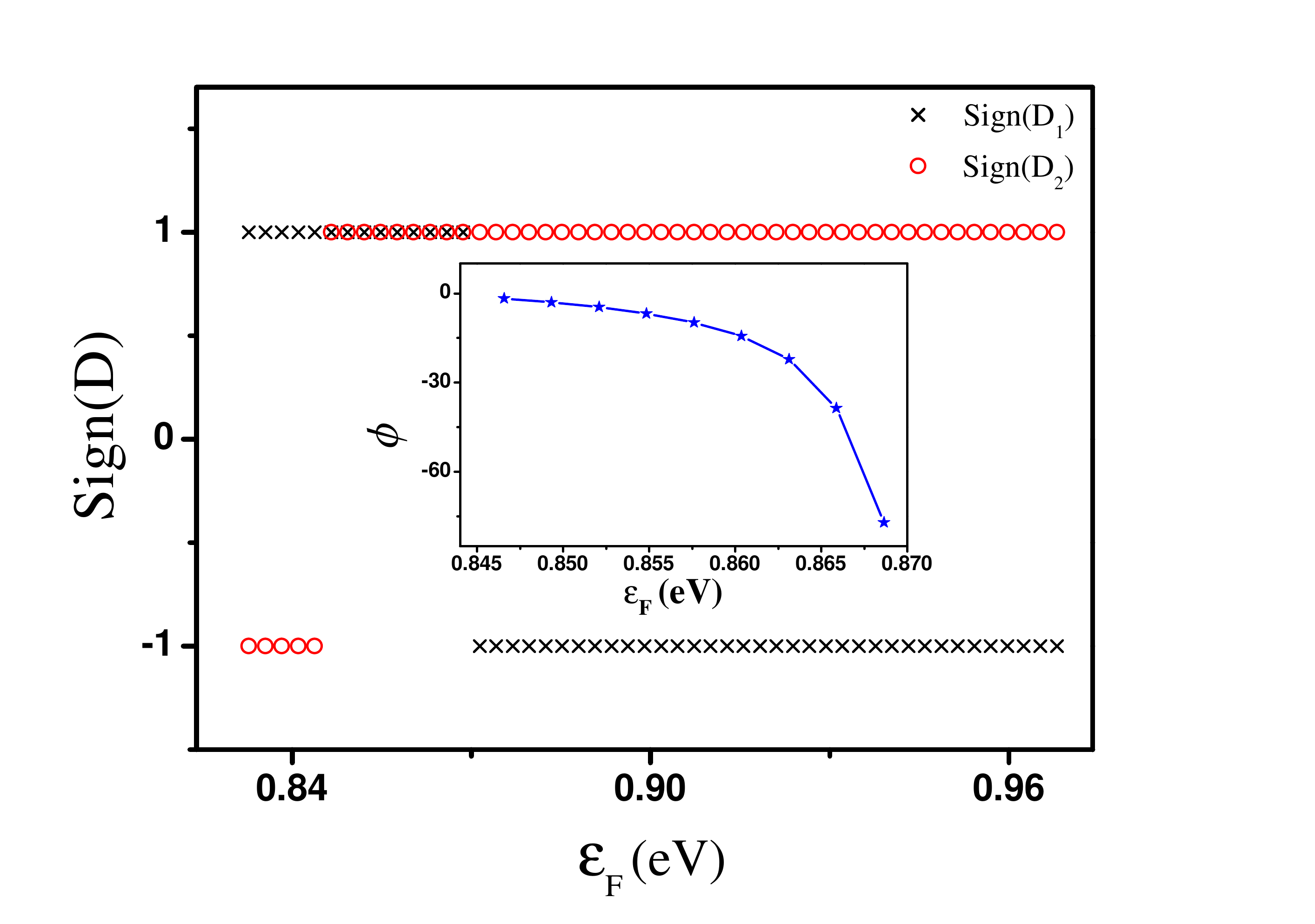}
\caption{ (Color online) Sign of $D_1$ and $D_2$ as a function of the Fermi energy in the conduction band for ${\bf R}=(4\sqrt{3}a, 0)$. Inset figure shows the angle $\phi=\tan^{-1}(J_{DM}/J_H)$ between two magnetic moments when $D_1$ and $D_2$ are both positive. \label{fig:Fig7}}
\end{figure}

Figure~\ref{fig:Fig7} shows the sign of $D_1$ and $D_2$ as a function of the Fermi energy in the conduction band when the distance vector between impurities is ${\bf R}=(4\sqrt{3}a, 0)$. A saddle point, minimum and maximum regions are taken place by increasing the Fermi energy from the bottom of conduction, respectively. The inset figure shows the angle $\phi$ between moments for the middle region in which both $D_1$ and $D_2$ are positive and $\theta=\pi/2$.

\section{Summary and Conclusions}\label{sect:conclusion}

We study the influence of spin-orbit coupling
on the RKKY interaction in a monolayer MoS$_2$ system. We use the
Green's function method in the continuum model.
We show that the rotation of the itinerant spin
causes a twisted interaction between
two magnetic adatoms which consists of different RKKY coupling terms,
the Heisenberg, Dzyaloshinsky-Moriya and Ising interactions. We explore different scenarios of two magnetic moments in terms of the different RKKY coupling values. We also find that
the interaction terms are very sensitive to the Fermi energy values and
 change dramatically from doped to undoped systems. For undoped system, we find a short-range interaction between the moments with most probably
an antiferromagnetic-like structure. The dependence of the interaction on the distance
$R$ between two local magnetic moments is
found to be $R^{-2}$ in a doped
monolayer MoS$_2$ sheet. We obtain that for a dilute electron doping, the Dzyaloshinsky-Moriya
interaction can be larger than the Heisenberg interaction for certain value of $R$. A beating pattern of oscillations of the RKKY interaction which occurs for the doped system is discussed.

We conclude that the monolayer MoS$_2$ is a quite good platform to explore different types of spin Hamiltonian. The interaction terms in the single layer MoS$_2$ are quite similar to those terms appearing in the topological insulator system~\cite{ref:TI}, however the interaction terms here are direction dependent where in the topological insulators the interactions are only distance dependent.

\section{Acknowledgments}

This work was partially supported by IPM grant.

\end{document}